Accepted to **ADVANCED MATERIALS**

# A Crown-Ether Loop-Derivatized Oligothiophene Doubly Attached on Gold Surface as Cation-Binding Switchable Molecular Junction

By *Truong Khoa Tran, Kacem Smaali, Marie Hardouin, Quentin Bricaud, Maïténa Oçafrain, Philippe Blanchard,\* Stéphane Lenfant,\* Sylvie Godey, Jean Roncali,\* and Dominique Vuillaume\**

Dr. T.K. Tran, Dr. M. Hardouin, Dr. Q. Bricaud, Dr. M. Oçafrain, Dr. P. Blanchard, Dr. J. Roncali
LUNAM Université, Université d'Angers, CNRS UMR 6200, MOLTECH-Anjou, Group Linear Conjugated Systems, 2 Bd Lavoisier,
49045 Angers cedex (France)
E-mail: Philippe.Blanchard@univ-angers.fr, Jean.Roncali@univ-angers.fr

Dr. K. Smaali, Dr. S. Lenfant, Dr. S. Godey, Dr. D. Vuillaume
Institute for Electronics Microelectronics & Nanotechnology, CNRS, University of Sciences and Technologies of Lille, BP60069, Avenue Poincaré,
59652 Villeneuve d'Ascq (France)
E-mail: stephane.lenfant@iemn.univ-lille1.fr, dominique.vuillaume@iemn.univ-lille1.fr

Molecular junctions and switches are a focus of considerable current interest as possible basic components of future molecular electronic devices.[1,2]

During the past decades molecular switches based on different concepts and mechanisms have been proposed, synthesized and investigated.[3] These include reversible bond breaking,[1,4,5] translocation of part of multi-components molecular assemblies,[6,7] proton transfer,[1,4,8] or reversible conformational changes of a system with geometry-dependent electronic properties.[9-11]

The interplay of the cation-binding properties of crown-ether and conformational changes has already been investigated. Thus, Shinkai used the photoisomerization of azobenzene units inserted in crown ether systems to modulate the cation binding properties of the cavity.[12] From a different viewpoint, we have shown that the cation-binding ability of a polyether loop attached at two fixed points of an oligothiophene chain can serve as driving force to generate changes in the geometry and hence electronic properties of the conjugated system.[13] A similar approach has been previously used for the synthesis of polythiophene-based sensors[14] and more recently for the control of the intramolecular photoinduced electron transfer in crown ether bridged oligothiophenes.[15]

In this context, we now report on the surface immobilization by double fixation on gold surface of a dithiol quaterthiophene **1-SH** derivatized with a polyether loop (Scheme 1). In recent years we have extensively studied the structural conditions and synthetic approaches allowing the horizontal double fixation of conjugated oligothiohenes as monolayers on gold surface.[16] After an analysis of the cation complexation properties of the related acetyl-protected dithiol molecule **1** in solution by UV-Vis spectroscopy and cyclic voltammetry, the preparation of monolayers by double fixation of the oligothiophene chain on gold surface will be described, the structure properties of these monolayers will be investigated using cyclic voltammetry, ellipsometry, water contact angle measurement, XPS and their electrical properties will be assessed by contacting the monolayer with a conducting eutectic GaIn drop. Finally preliminary results on the cation binding properties of the monolayer and investigations on the use of the immobilized molecule as a switchable molecular junction will be presented and possible transport mechanisms will be discussed.

## Scheme 1

Compound **1** has been synthesized by deprotection/functionalization of the appropriately protected thiolate groups according to the already published method.[16b] The detailed synthesis of this molecule and of some parent compounds will be reported elsewhere.[17] The identity and purity of compound **1** were established by $^1$H and $^{13}$C NMR spectrometry and HR mass spectrometry giving satisfactory results (see Supporting Information). Dithiol **1-SH** was prepared by reduction of the thioester groups of **1** using DIBAl-H. Monolayers were elaborated just after purification of **1-SH** by chromatography on silica gel to avoid formation of di- or polydisulfide oligomers.

**Metal complexation in solution.** The metal cation complexing properties of compound **1** have been analyzed by UV-Vis spectroscopy and cyclic voltammetry. Preliminary tests using UV-Vis absorption spectroscopy in the presence of various metal cations ($Li^+$, $Na^+$, $Cs^+$, $Ba^{2+}$, $Sr^{2+}$, $Cd^{2+}$ and $Pb^{2+}$) have shown that only $Pb^{2+}$ is complexed by compound **1**.

Figure 1a shows the UV-Vis absorption spectrum of **1** in $CH_3CN$ upon stepwise addition of substoichiometric amounts of $Pb^{2+}$. The initial spectrum shows two maxima at 279 and 380 nm. Addition of $Pb^{2+}$ produces an intensification of the two absorption bands. These changes occur around an isobestic point at 398 nm which suggests interconversion between two species. Addition of one equivalent of $Pb^{2+}$ leads to a 5 nm hypsochromic shift of the 380 nm maximum. This blue-shift indicative of a decrease of $\pi$-conjugation of the quaterthiophene skeleton can be ascribed to an increase of the twist angle of the central C(5')-C(2'') bond. As shown in Fig. 1a, the blue-shift stops and the intensity of the two bands remains constant after addition of more than one equivalent of $Pb^{2+}$ suggesting the formation of a 1:1 complex with $Pb^{2+}$.

**Figure 1**

Figure 1b shows the cyclic voltammogram (CV) of **1** upon complexation of $Pb^{2+}$. The initial CV of **1** exhibits two **one-electron** reversible oxidation waves corresponding to the successive formation of **the radical-cation** ($E_{pa}^1 = 0.86$ V) and dication ($E_{pa}^2 = 0.99$ V) of the **quaterthiophene backbone**. The addition of increasing amounts of $Pb^{2+}$ to a solution of **1** produces a decrease in intensity of the first oxidation peak which completely disappears after addition of one equivalent of $Pb^{2+}$. The CV then exhibits a single two-electron reversible oxidation peak at 1.02 V. Thus, the first and second initial oxidation peaks undergo a 160 mV and 30 mV positive shifts, respectively. This result can be related to the existence of repulsive Coulombic interactions between $Pb^{2+}$ and the radical cation or the dication of the quaterthiophene skeleton, and also to a decrease of the *+M* mesomeric effect of the two sulfur atoms in 4' and 3'' positions due to the participation of their free electron pairs to the complexation of $Pb^{2+}$.[18] In addition, no more change is observed beyond one equivalent of $Pb^{2+}$ which further supports the formation of a 1:1 complex between **1** and $Pb^{2+}$.

This electrochemical behavior is similar to that of many redox active crown ethers,ref where the major changes observed in the first CV wave corresponds to the complexation of the métal cation while the quasi-invariance of the potential of the second oxidation waves is related to the expulsion of the cation by repulsive electrostatic interactions between the cation and the positively charged oligothiophene. In our case, the situation is somehow complicated by the fact that the oligothiophene chain undergoes geometrical changes (and hence modification of its redox properties) during the complexation/decomplexation processes (see ref 9 for detailed discussions on these points).

**Preparation, characterization and metal complexation properties of monolayers of 1-SH.**
Monolayers of **1-SH** were prepared under a controlled argon atmosphere by immersion of gold bead electrodes[16] during 48 h in a millimolar solution of compound **1-SH** in $CH_2Cl_2$. The resulting electrodes were then rinsed with $CH_3CN$ before CV analysis. Reproducible monolayers were obtained when dithiol **1-SH** was freshly generated and chromatographed to prevent the possible formation of multilayers of disulfide compounds.[16,19]

The CV of monolayer of **1-SH** exhibits only one broad reversible oxidation peak at 1.05 V/SCE. As already observed for a parent 4T derivative containing ferrocene,[16b] this single oxidation wave includes in fact the two successive one-electron oxidation steps of 4T. The occurrence of a single oxidation wave suggests a broadening of the individual waves due to intermolecular interactions as observed in the CV of oligo- and polythiophenes.[20] The intensity of peak currents varies linearly with scan rate as expected for a surface-confined electrochemical reaction. These monolayers exhibit a good electrochemical stability and no variation of the CV was observed after 200 cycles between 0.50 and 1.10 V/SCE at 1 Vs⁻¹. The surface coverage $\Gamma$ of molecules **1-SH** has been determined by using gold planar electrode of 0.2 cm$^2$ and integration of the voltammetric peak after correction for double layer charge.[21] A typical value of 1 x 10$^{-10}$ mol cm$^{-2}$ (i.e. 0.6 x 10$^{14}$ molecules/cm$^2$) was obtained for $\Gamma$, in agreement with the formation of a monolayer on gold.

The complexation properties of $Pb^{2+}$ by monolayers of **1-SH** have been analyzed by cyclic voltammetry. The CV of monolayers of **1-SH** have been recorded repeatedly and alternately in a solution of 0.1 M $Bu_4NPF_6/CH_3CN$ in the absence or presence of $Pb(ClO_4)_2$ (5 mM) (Figure 1c). A significant positive shift of 90 mV of the oxidation wave is observed for each CV cycle recorded in the presence of $Pb^{2+}$. This result shows that the monolayer of **1-SH** is sensitive to the presence of $Pb^{2+}$ as expected from the electrochemical behavior observed in solution.

Monolayers of **1-SH** have been also prepared on evaporated gold layer (200 nm) on silicon wafers recovered by a 10 nm adhesion layer of titanium. After monolayer fixation, the gold substrates were thoroughly rinsed with acetonitrile prior to further experiments. Thickness measurement of the monolayers by ellipsometry, gave a values of 14 ± 2 Å. This value is in good agreement with a monolayer formation and with the height of 18 Å calculated for **1-SH** in a geometry expected for a double fixation, as deduced from MOPAC 3D software and assuming a S-Au distance of about 2 Å (see Supporting Information).[22]

The relatively low values of water contact angle for monolayers of **1-SH** ($\theta_{H2O}$ = 74 ± 2°) show that these surfaces present a hydrophilic character consistent with the presence of the oxyethylene chains on the top of the surface.[23]

The water contact angles and monolayer thicknesses are not much affected by $Pb^{2+}$ complexation. For a monolayer of **1-SH** immersed in a 5 mM solution $Pb(ClO_4)_2.3H_2O$ in $CH_3CN$ for 48 h and rinsed with $CH_3CN$, we have measured $\theta_{H2O}$ = 71 ± 2° and a thickness of 16 ± 2 Å. This feature indicates that the integrity of the monolayer is globally preserved upon $Pb^{2+}$ complexation.

The survey of XPS spectrum (not shown here) of the **1-SH** monolayer shows the presence of the different atoms of the molecule: carbon (peak C1s), oxygen (peak O1s), sulfur (peaks S2s and S2p) and lead (peak Pb4f) after complexion. The C1s spectrum shows (Fig. 2a) two components at 285.5eV (main peak) and 287 eV (shoulder). The main peak is attributed to C–C in good agreement with results published for monolayers grafted on gold.[24] The shoulder peak is attributed to C-O,[25] with an experimental area peak ratio C-C/C-O of 3.3 (expected value of 26/8=3.25). The S2s and O1s peaks appear at 228.4 eV and 532.9 eV, respectively, without significant variation after $Pb^{2+}$ complexation.

**Figure 2**

The high-resolution XPS spectrum of the S2p region of monolayers of **1-SH** shows two doublets (Fig. 2b). The peaks at 162.5 eV and 163.7 eV of the lower energy doublet S2p$_{3/2,1/2}$ of monolayers may be assigned to the thiol chemisorbed on the Au surface (S-Au bond).[24a,26] The other doublet S2p$_{3/2,1/2}$ at 164.3 and 165.5 eV corresponds to the other sulfur atoms of compound **1-SH** (S-C bond). Comparison of the areas of these two doublet signals leads to an estimated S-C/S-Au ratio of 3.8 ± 1 which is close to the theoretical value expected for a double fixation (S-C/S-Au = 4). Thus, XPS results confirm that the majority of molecules **1-SH** are doubly fixed on the gold surface.

After $Pb^{2+}$ complexation by the **1-SH** monolayer, the energy values of the four components of the S2p peak are not significantly modified (less than 0.2 eV). In addition, we also observed (Fig. 2c) the presence of Pb4f peaks at 139.8 eV and 144.9 eV for the doublet Pb4f$_{7/2,5/2}$.[27] The Pb/C ratio of 4 x $10^{-3}$ determined by XPS shows that around 13% of $Pb^{2+}$ is still present in the monolayer.

**Electronic transport properties.** Current-voltage (I-V) are measured on Au-molecule-metal junctions by contacting the top surface of the monolayer with an eutectic Gallium-Indium (eGaIn) drop (see Experimental section).[28] Typical I-V curves are shown in Fig. 3a before and after complexation of $Pb^{2+}$. For the pristine SAM, a very low current (slightly higher than the sensitivity of our amp-meter of a ca. 0.01 pA) is measured between -1 V to + 1 V. The I-V curves display a rather symmetrical behavior for negative and positive voltages (i.e. the ratio $I_{V<0}/I_{V>0}$ is always around 1). Applying the transition voltage spectroscopy (TVS) technique,[29] i.e. plotting the same data as $\ln(I/V^2)$ *vs.* 1/V (Fig. 3b),[30] we get a positive and negative transition voltage $V_T$ (i.e. the voltage at which $\ln(I/V^2)$ is minimum) at +1.09 ± 0.02 V and **-1.11** ± 0.02 V. The SAMs with $Pb^{2+}$ show a rather different behavior, with a huge increase of the current at low bias (a ratio up to 280 at around 1.15 V and 1.6 x $10^3$ at -1.2 V). We also note a slight asymmetrical behavior with more current at negative bias than at positive bias (with a max ratio of about 9). Then, at higher bias (< -2V and > 2V) the two I-V curves overlap. TVS (Fig. 3b) show a decrease of $V_T$, which are -0.52 ± 0.04 V and +0.86 ± 0.04 V, these asymmetric values (in absolute values) reflecting the asymmetrical I-V behavior already mentioned.[31] Recent discussions in the literature pointed out how it can be possible to relate $V_T$ with the energy position of the LUMO or HOMO levels in the junctions and with the voltage division factor γ, a parameter that describes the degree of symmetry or asymmetry of the molecular orbitals in the junction (-0.5 ≤ γ ≤ 0.5, γ = 0 being the case of a symmetrical coupling of the molecular orbitals between the two electrodes).[30,31] Following the analytical model of Bâldea[31b] (see supporting information), we determine the energy level $\varepsilon_0$ of the molecular orbital involved in the electrical transport (with respect to the Fermi energy of the electrodes), and γ, directly from the measured $V_{T+}$ and $V_{T-}$. For the pristine **1-SH** junction, we get $\varepsilon_0$ = 0.96 eV, and γ = 0. After $Pb^{2+}$ complexation, $\varepsilon_0$ decreases to 0.56 eV and γ = ± 0.1 depending whether $\varepsilon_0$ corresponds to the LUMO or the HOMO, respectively. Based solely on these I-V measurements and analytical model, we can conclude that the $Pb^{2+}$ complexation induces a decrease of the energy offset between the Fermi energy of the electrodes and the molecular orbital involved in the electron transport, as well as, an asymmetry in the localization of this orbital and/or in the molecule-electrode coupling. From the TVS analysis, we propose a possible energy diagram of the junction with and without $Pb^{2+}$ (see supplementary information). However, from this simple analysis it is not possible to discriminate which molecular orbital is involved in the transport. A more detailed analysis and a comparison with ab-initio calculations are in progress and will be reported elsewhere.[17]

In conclusion, we clearly demonstrate that the $Pb^{2+}$ complexation by a **1-SH** molecule moves one of the molecular orbitals close to the Fermi energy of the electrodes, inducing a significant increase (up to $1.6 \times 10^3$) of the current at low bias, while only about one $Pb^{2+}$ atom is captured per 7-8 molecules. The distortion of the molecular conformation after $Pb^{2+}$ complexation also leads to an asymmetrical molecular coupling in the junction with the two electrodes as evidenced by the rectification behavior.

*Experimental*

Gold substrates consist of 200 nm Au film thickness, evaporated onto silicon wafers covered by titanium or chromium sublayer (10 nm) deposited under ultrahigh vacuum. Electrochemical experiments were carried out with a PAR 273 potentiostat-galvanostat in a three-electrode single compartment cell. We measured the water contact angle with a remote-computer controlled goniometer system (DIGIDROP by GBX, France). We recorded spectroscopic ellipsometry data in the visible range using an UVISEL (Jobin Yvon Horiba) Spectroscopic Ellipsometer equipped with a DeltaPsi 2 data analysis software. XPS measurements were performed with a Physical Electronics 5600 spectrometer fitted in an UHV chamber with a residual pressure of $2 \times 10^{-10}$ Torr. We performed current-voltage measurements by Eutectic

GaIn drop contact (eGaIn 99.99%, Ga:In; 75.5:24.5 wt% from Alfa Aesar). I-V curves were acquired with an Agilent semiconductor parameter analyzer 4156C with an ultimate sensitivity of 0.01 pA. More details in the Supporting Information.

*Acknowledgements*
The Ministère de la Recherche is acknowledged for the PhD grant of T.K. Tran. This work was financially supported by ANR-PNANO (OPTOSAM project ANR-06-NANO-016) and the CNANO Nord-Ouest. The authors thank O. Alévêque (MOLTECH-Anjou) for technical assistance and the PIAM of the University of Angers for the characterization of organic compounds.

Supporting Information is available online from Wiley InterScience or from the author.

Received: ((will be filled in by the editorial staff))
Revised: ((will be filled in by the editorial staff))
Published online: ((will be filled in by the editorial staff))

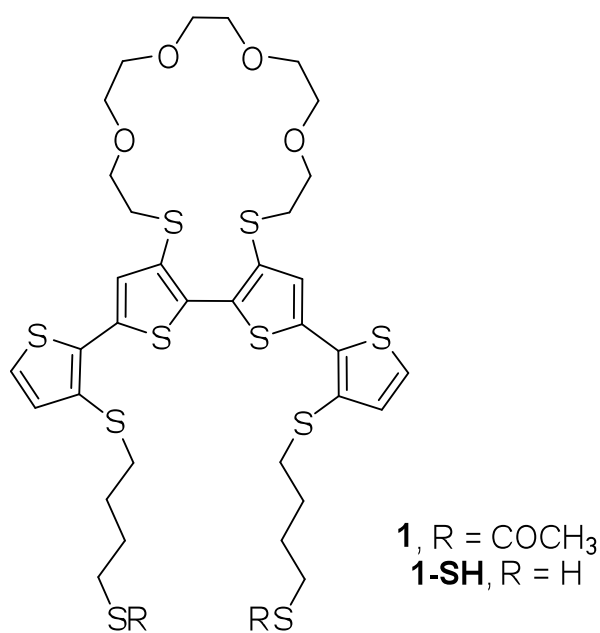

**1**, R = COCH₃
**1-SH**, R = H

**Scheme 1.** Crown-ether quaterthiophenes **1** and **1-SH**.

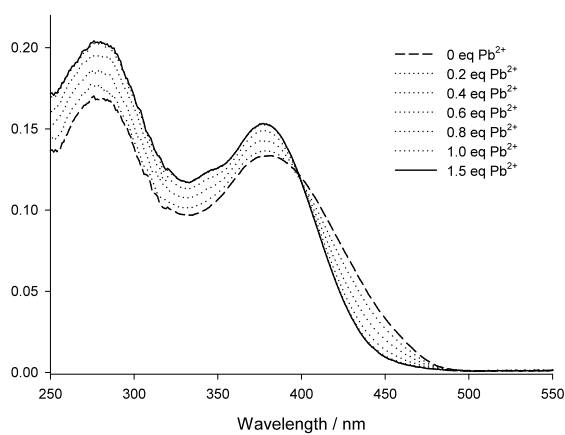

(a)

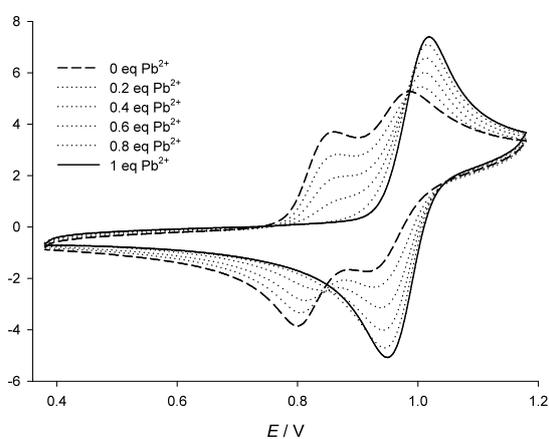

(b)

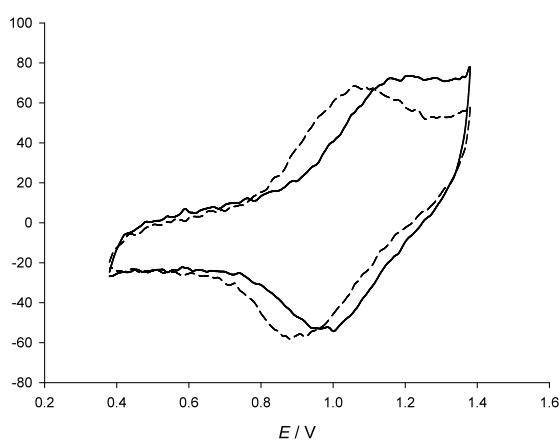

(c)

**Figure 1.** (a) Changes in the UV-vis spectrum of **1** (1 x $10^{-5}$ M) in CH$_3$CN *vs* number of Pb$^{2+}$ eq. added as perchlorate. (b) CV of **1** (0.5 mM in 0.10 M Bu$_4$NPF$_6$/CH$_3$CN, scan rate 0.1 V s$^{-1}$, Pt working electrode, ref. SCE) in the presence of increasing amounts of Pb$^{2+}$ added as perchlorate by increments of 0.2 equivalent. (c) CVs of a monolayers of **1-SH** on a gold bead electrode in 0.10 M Bu$_4$NPF$_6$/CH$_3$CN in the absence (dotted line) and in the presence (solid line) of Pb(ClO$_4$)$_2$.3H$_2$O (5 mM), scan rate 20 V s$^{-1}$, ref. SCE.

(a)

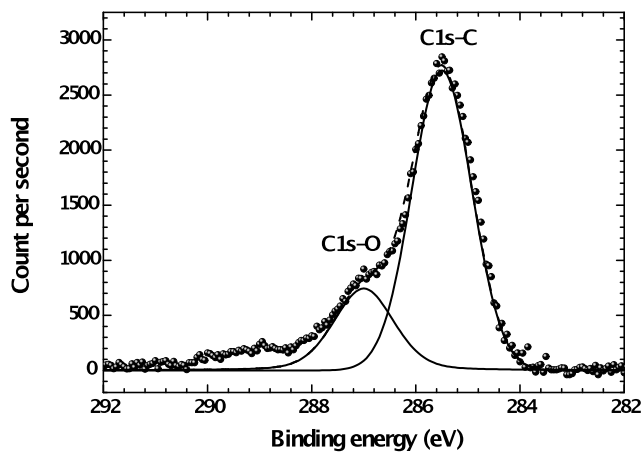

(b)

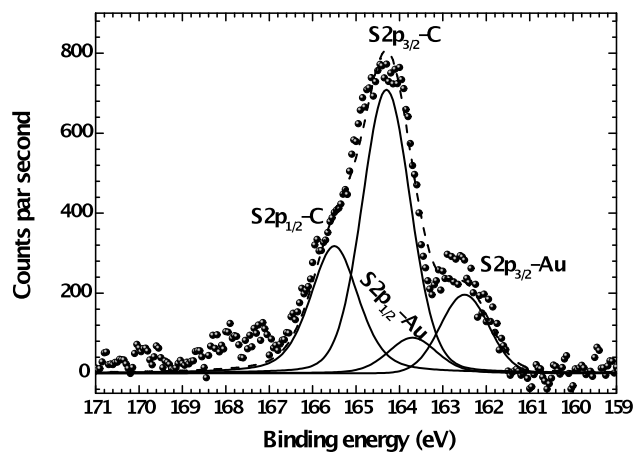

(c)

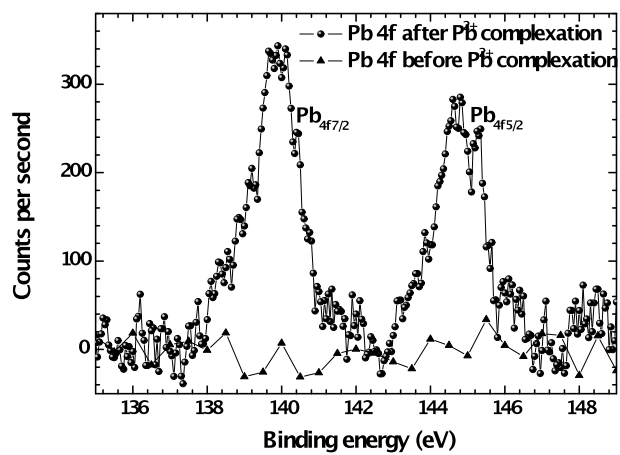

**Figure 2.** (a) C1s region of the XPS spectrum of 1-SH monolayer before exposure to $Pb^{2+}$. (b) S2p region of the XPS spectrum of 1-SH monolayer before exposure to $Pb^{2+}$. (c) XPS spectra of the Pb4f region before and after $Pb^{2+}$ complexation.

(a)

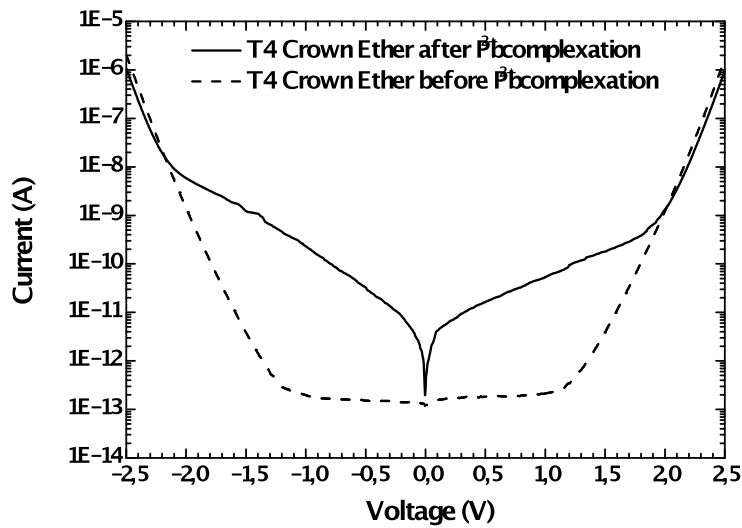

(b)

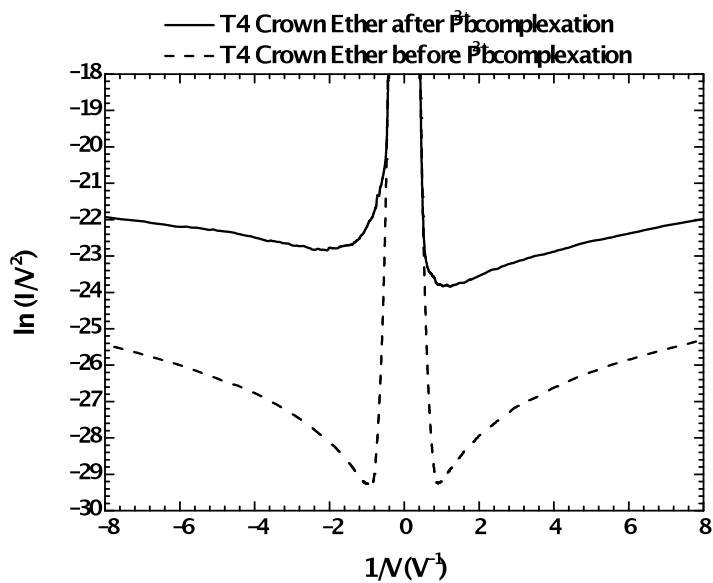

**Figure 3.** Typical I-V curves (averaged over 5 measurements) for the Au/molecules/eGaIn junction before and after Pb$^{2+}$ complexation. (b) TVS plots of the same data as in Fig. 3a.

A **crown-ether dithiol quaterthiophene** is synthesized for its covalent immobilization by double fixation on gold surface. While the corresponding dithioester precursor exhibits $Pb^{2+}$ complexation properties in solution as evidenced by UV-vis spectroscopy and cyclic voltammetry, this $Pb^{2+}$ affinity is still maintained for monolayers of the dithiol on gold. In addition, current-voltage measurements by Eutectic GaIn drop contact on the monolayer show a significant increase (up to $1.6 \times 10^3$) of the current at low bias after $Pb^{2+}$ complexation.

Truong Khoa Tran, Kacem Smaali, Marie Hardouin, Quentin Bricaud, Maïténa Oçafrain, Philippe Blanchard, Stéphane Lenfant, Sylvie Godey, Jean Roncali, and Dominique Vuillaume

A Crown-Ether Loop-Derivatized Oligothiophene Doubly Attached on Gold Surface as Cation-Binding Switchable Molecular Junction

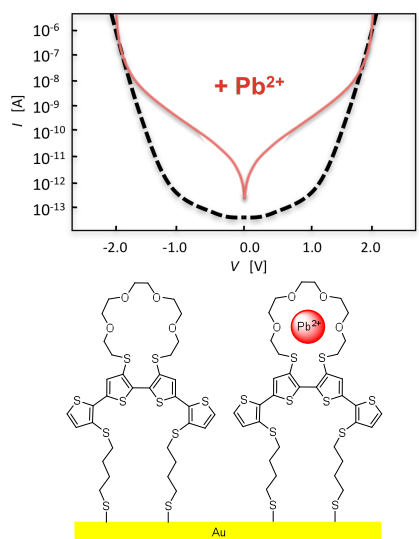

# A Crown-Ether Loop-Derivatized Oligothiophene Doubly Attached on Gold Surface as Cation-Binding Switchable Molecular Junction

T.K. Tran, M. Hardouin, Q. Bricaud, M. Oçafrain, P. Blanchard, J. Roncali
LUNAM Université, Université d'Angers, CNRS UMR 6200, MOLTECH-Anjou, Group
Linear Conjugated Systems, 2 Bd Lavoisier,
49045 Angers cedex (France)

K. Smaali, S. Lenfant, S. Godey, D. Vuillaume
Institute for Electronics Microelectronics & Nanotechnology, CNRS, University of Sciences
and Technologies of Lille, BP60069, Avenue Poincaré,
59652 Villeneuve d'Ascq (France)

# Supporting Information

**3,3'''-Bis(6-oxo-5-thiaheptylsulfanyl)-4',3''-bis(1,16-dithia-4,7,10,13-tetraoxahexadecane-1,16-diyl)-2,2':5',2'':5'',2'''-quaterthiophene (1).**

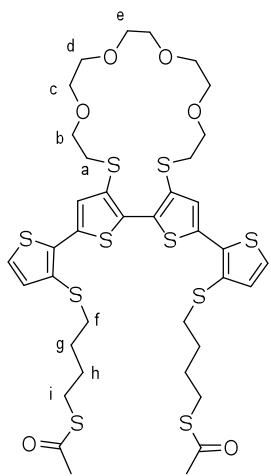

Yellow oil. $^1H$ NMR (500 MHz, CDCl$_3$, δ): 7.38 (s, 2H, H$_{thio}$), 7.22 (d, 2H, $^3J$=5.2 Hz, H$_{thio}$), 7.03 (d, 2H, $^3J$=5.2 Hz, H$_{thio}$), 3.67 (t, 4H, $^3J$=6.75 Hz, H$_b$), 3.60 and 3.55 (2s, 12H, H$_c$, H$_d$, H$_e$), 3.02 (t, 4H, $^3J$=6.75 Hz, H$_a$), 2.88-2.83 (m, 8H, H$_f$, H$_i$), 2.31 (s, 6H, CH$_3$-CO), 1.71-1.65 (m, 8H, H$_g$, H$_h$); $^{13}C$ NMR (75 MHz, CDCl$_3$, δ): 195.8, 136.0, 135.4, 132.7, 132.4, 130.9, 129.6, 128.1, 123.8, 70.73, 70.70, 70.42, 70.05, 35.6, 35.2, 30.7, 28.55, 28.52, 28.50; IR (NaCl): ν = 1687 (C=O), 1118 cm$^{-1}$ (C-O-C); UV-vis (CH$_2$Cl$_2$): λ$_{max}$ (log ε) = 279 (4.23), 380 nm (4.15); MALDI MS: 920 [M$^{+\bullet}$], 943 [M+Na$^+$], 959 [M+K$^+$]. HRMS (ESI, m/z): calcd for C$_{38}$H$_{48}$O$_6$S$_{10}$Na 943.0556, found, 943.0538. HRMS (ESI, m/z): calcd for C$_{38}$H$_{48}$O$_6$S$_{10}$K 959.0295, found 959.0319.

**3,3'''-Bis(4-mercaptobutylsulfanyl)-4',3''-bis(1,16-dithia-4,7,10,13-tetraoxahexadecane-1,16-diyl)-2,2':5',2'':5'',2'''-quaterthiophene (1-SH).**

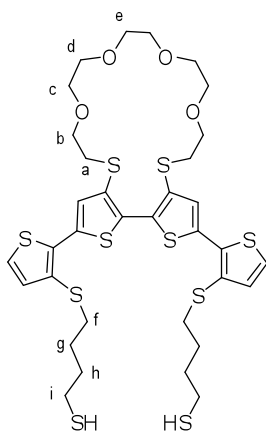

Yellow oil. $^1$H NMR (500 MHz, CDCl$_3$, δ): 7.38 (s, 2H, H$_{thio}$), 7.22 (d, 2H, $^3J$ = 5.2 Hz, H$_{thio}$), 7.05 (d, 2H, $^3J$ = 5.2 Hz, H$_{thio}$), 3.66 (t, 4H, $^3J$ = 6.8 Hz, H$_b$), 3.58 (m, 12H, H$_c$, H$_d$, H$_e$), 3.02 (t, 4H, $^3J$ = 6.8 Hz, H$_a$), 2.80 (m, 4H, H$_f$), 2.50 (q, 4H, H$_i$), 1.72 (m, 8H, H$_g$, H$_h$), 1.32 (t, 2H, $^3J$ = 7.9 Hz, SH). UV-vis (CH$_2$Cl$_2$): λ$_{max}$ (log ε) = 295, 385 nm; MALDI MS: 874.7 [M$^{+\cdot}$ + K$^+$], 835.73 [M$^{+\cdot}$].

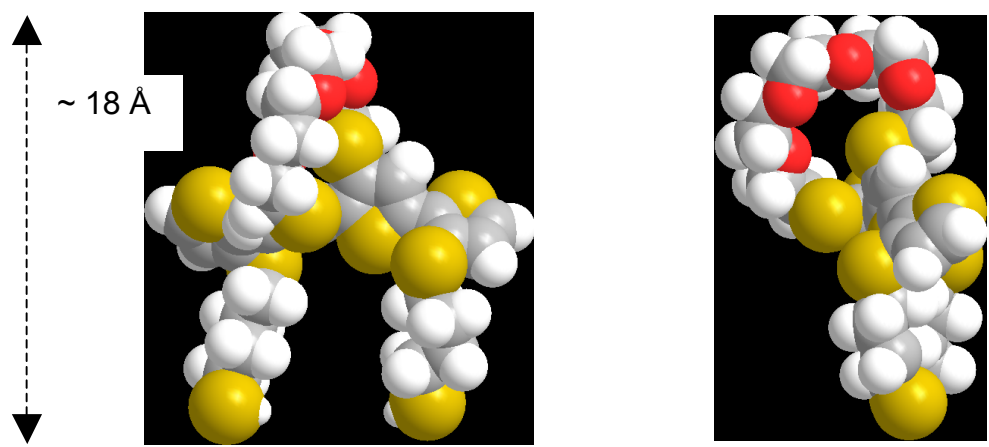

**Figure S1**. Molecular structure of dithiol **1-SH** drawn from ellipsometry data in a conformation leading to a double fixation on gold (from MOPAC-ChemDraw 3D optimization).

**Gold substrate fabrication:** Ellipsometry, contact angles and XPS analyses were performed on monolayers prepared from gold films of 200 nm thickness, evaporated onto silicon wafers covered by titanium or chromium sublayer (10 nm) deposited under ultrahigh vacuum. After gold deposition, annealing (rms roughness less than 0.88 nm determined on 2x2μm images) and cleaning in a HCl/HNO$_3$/H$_2$O$_2$ (3/1/16) mixture for 5 minutes led to flat gold terraces. Silicon and gold were respectively purchased from Siltronix and Goodfellow.

**Cyclic voltammetry:** Electrochemical experiments were carried out with a PAR 273 potentiostat-galvanostat in a three-electrode single compartment cell equipped with platinum or gold disk of 2 mm diameter and modified gold beads as working electrodes, a platinum wire counter electrode and a silver wire as pseudo-reference electrode. The ferricinium/ferrocenium couple was used as internal reference (E°(Fc$^+$/Fc) = 0.405 V/SCE in 0.1 M Bu$_4$NPF$_6$/CH$_3$CN or CH$_2$Cl$_2$). Potentials were then expressed towards a saturated calomel reference electrode (SCE).

**Contact-angle measurements**: We measured the water contact angle with a remote-computer controlled goniometer system (DIGIDROP by GBX, France). We deposited a drop (10-30 μL) of desionized water (18 MΩ.cm⁻¹) on the surface and the projected image was acquired and stored by the computer. Contact angles were extracted by contrast contour image analysis software. These angles were determined few seconds after application of the drop. These measurements were carried out in a clean room (ISO 6) where the relative humidity (50%) and the temperature (22°C) are controlled. The precision with these measurements are ± 2°.

**Spectroscopic ellipsometry**: We recorded spectroscopic ellipsometry data in the visible range using an UVISEL (Jobin Yvon Horiba) Spectroscopic Ellipsometer equipped with a DeltaPsi 2 data analysis software. The system acquired a spectrum ranging from 2 to 4.5 eV (corresponding to 300 to 750 nm) with intervals of 0.1 eV (or 15 nm). Data were taken at an angle of incidence of 70°, and the compensator was set at 45.0°. We fitted the data by a regression analysis to a film-on-substrate model as described by their thickness and their complex refractive indexes. First, we recorded a background before monolayer deposition for the gold coated substrate. Secondly, after the monolayer deposition, we used a 2 layers model (substrate/SAM) to fit the measured data and to determine the SAM thickness. We used the previously measured optical properties of the gold coated substrate (background), and we fixed the refractive index of the organic monolayer at 1.50. The usual values in the literature for the refractive index of organic monolayers are in the range 1.45-1.50.56, 57 We can notice that a change from 1.50 to 1.55 would result in less than 1 Å error for a thickness less than 30 Å. We estimated the accuracy of the SAM thickness measurements at ± 2 Å.

**XPS measurements**: We performed XPS measurements to control the chemical composition of the SAMs and to detect any contaminant. We used a Physical Electronics 5600 spectrometer fitted in an UHV chamber with a residual pressure of $2 \times 10^{-10}$ Torr. High resolution spectra were recorded with a monochromatic AlKα X-ray source (hν=1486.6 eV), a detection angle of 45° as referenced to the sample surface, an analyzer entrance slit width of 400mm and with an analyzer pass energy of 12 eV. In these conditions, the overall resolution as measured from the full-width half-maximum (FWHM) of the Ag 3d5/2 line is 0.55 eV. Semi-quantitative analysis were completed after standard background substraction according to Shirley's method.[1] Peaks were decomposed by using Voigt functions and a least-square minimization procedure and by keeping constant the Gaussian and Lorentzian broadenings for each component of a given peak.

**Electrical measurements:** We performed current-voltage measurements by Eutectic GaIn drop contact (eGaIn 99.99%, Ga:In; 75.5:24.5 wt% from Alfa Aesar). We used a method close to the one developed by Chiechi et al.[2] We formed a drop of eGaIn at the extremity of a needle fixed on a micromanipulator. By displacing the needle, we brought the drop into contact with a sacrificial surface, and we retracted the needle slowly. By this technique, we formed a conical tip of eGaIn with diameter ranging from 50 to 200 μm (corresponding to contact area ranging from $10^{-5}$ to $10^{-3}$ cm²). This conical tip was then put into contact with SAM (under control with a digital video camera). Voltage was applied on the eGaIn drop. For each SAM, we measured 5 I-V curves by moving the same eGaIn drop at different locations on the SAMs, thus all I-V curves are directly comparable (same contact area). We have a good reproducibility (Fig. S2). I-V curves were acquired with an Agilent semiconductor parameter analyzer 4156C with an ultimate sensitivity of 0.01 pA.

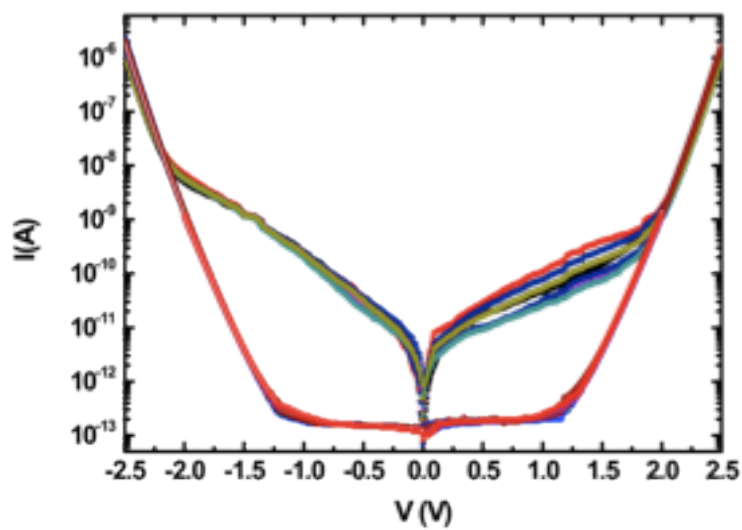

**Figure S2.** Reproducibility of the I-V measurements taken at 5 different locations for both the pristine monolayers and after $Pb^{2+}$ complexation.

**Transition Voltage Spectroscopy and molecular orbital energy diagram**. We follow the analytical model of Bâldea[3] to determine the energy level $\varepsilon_0$ of the molecular orbital involved in the electrical transport (with respect to the Fermi energy of the electrodes), and γ, directly

from the measured $V_{T+}$ and $V_{T-}$ according to:

$$|\varepsilon_0| = 2 \frac{e|V_{T+}V_{T-}|}{\sqrt{V_{T+}^2 + 10|V_{T+}V_{T-}|/3 + V_{T-}^2}}$$

$$\gamma = \frac{sign\ \varepsilon_0}{2} \frac{V_{T+} + V_{T-}}{\sqrt{V_{T+}^2 + 10|V_{T+}V_{T-}|/3 + V_{T-}^2}}$$

(S1)

where *sign ε₀* is + LUMO (- for HOMO). The voltage division factor γ is a parameter that

describes the degree of symmetry or asymmetry of the molecular orbitals in the junction (-0.5

≤ γ ≤ 0.5, γ = 0 being the case of a symmetrical coupling of the molecular orbitals between the

two electrodes).[4–6] This analytical model is based on a coherent transport model involving

one single orbital energy level. It has been validated against previous experiments on simple

"model" molecular junctions made with anthracene and terphenyl-based molecules.[7,8] In particular, the model of Bâldea reproduces correctly the full experimental current-voltage curves[3] using the two parameters ε₀ and γ deduced from Eq. S1.

Here, for the pristine SAM, with $V_{T+}$= 1.09 ± 0.02 eV and $V_{T-}$= -1.11 ± 0.02 eV, we

get ε₀ = 0.96 eV and γ = 0 (as expected for a symmetrical I-V curve). For the SAM with the

$Pb^{2+}$ complexation, we measured $V_{T+}$= 0.86 ± 0.04 eV and $V_{T-}$= -0.52 ± 0.04 eV, leading to ε₀

= 0.56 eV and γ = ± 0.1 depending whether ε₀ corresponds to the LUMO or the HOMO,

respectively.

From these parameters, we can propose an energy diagram of the molecular junction

without and with the $Pb^{2+}$ complexation. First, for the pristine SAM, a symmetric I-V and γ =

0 means that the molecular orbital involved in the electrical transport is geometrically at the center of the junction, *i.e.* at about the same distance from the two electrodes, and/or that the electronic coupling between this molecular orbital and the wave functions of the electrons in the metal electrodes are about of the same strength. Such a situation is schematically depicted in Fig. S3-a (red molecular orbitals) if we consider the LUMO as the orbital involved in the transport and in Fig. S3-b for the HOMO. A symmetrical geometry position of the LUMO and HOMO is consistent with the fact that both the HOMO and LUMO orbitals are localized on the quaterthiophene part of the molecule (the more conjugated part), and this part is about at the center of the molecule (see Fig. S1) since the short alkyl chain and the crown-ether loop have about the same size. For the SAM with $Pb^{2+}$, the energy diagrams are modified as indicated by the red arrows in Fig. S3-a and S3-b. First, the energy level (green lines) moves closer to the electrode Fermi energy. This reduction of the energy offset between the molecular orbital and the Fermi energy of the electrodes is in agreement with the experimental observation of the increase in the current after $Pb^{2+}$ complexation. In agreement with the higher current when a negative bias is applied on the GaIn electrode (rectification effect, see main text Fig. 3-a), the geometrical position also moves a little bit towards the Au electrode if the electron transport is mediated by the LUMO ($\gamma = 0.1$), or towards the GaIn electrode if it is the HOMO ($\gamma = -0.1$). A modification of the LUMO and HOMO localization, as well as a modification of their energy levels, after the complexation with $Pb^{2+}$ is likely, as the results of the interactions between Pb and the O and S atoms of the crown-ether loop. However, at this stage, and solely based on the experimental I-V measurements and this simple analysis it is not possible to discriminate which orbital (HOMO or LUMO) is involved in the transport. This point deserves more experiments (e.g. UPS and IPES measurements to directly determine the HOMO and LUMO positions, respectively), and a comparison with ab-initio calculations (in progress in our group).

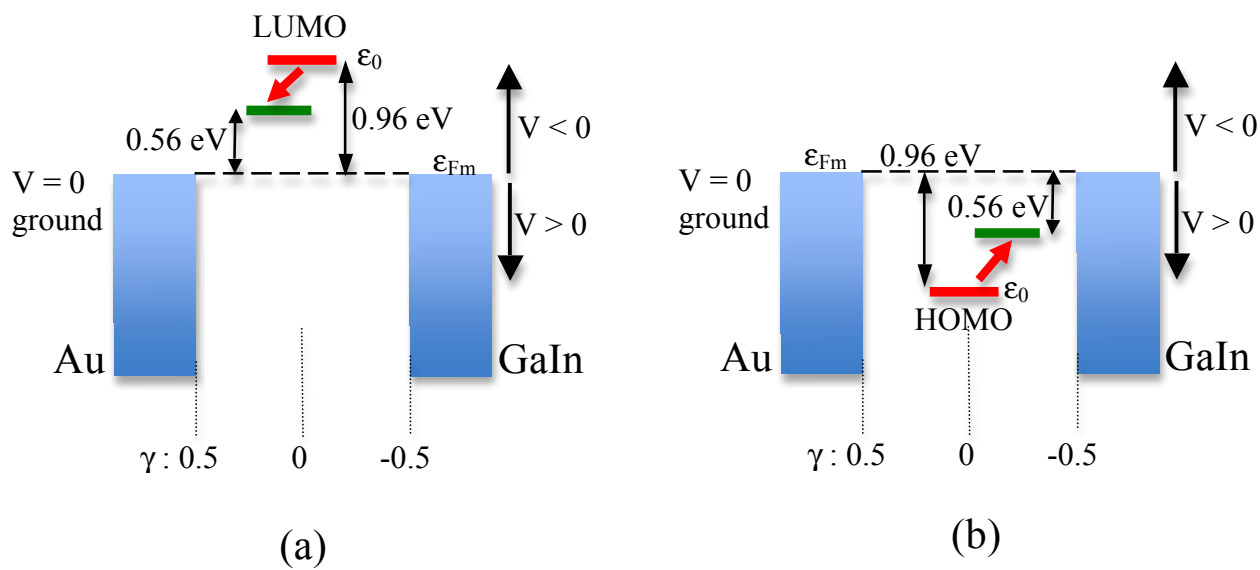

**Figure S3**. Schematic energy diagrams (at zero bias) of the molecular junctions without (red molecular orbitals) and with (green molecular orbitals) the $Pb^{2+}$: (a) with the hypothesis that the electron transport is LUMO-mediated, (b) with the hypothesis that the electron transport if mediated by the HOMO. The voltage division factor $\gamma$ is indicated, and varies from 0.5 to -0.5 (see text), 0 being the case of a symmetric position of the molecular orbital (or more generally speaking of the symmetric electronic coupling with the two electrodes). The Au electrode is grounded; the bias is applied on the GaIn electrode. $\varepsilon_{Fm}$ is the Fermi level of the electrodes and $\varepsilon_0$ the energy position of the molecular orbital.